\documentclass[conference,a4paper]{APSIPA2020}
\usepackage{amsmath,graphicx}
\usepackage{epstopdf}
\usepackage{color, soul}
\usepackage{url}
\usepackage{cite}
\usepackage{marvosym}
\usepackage{cite}
\usepackage{multirow}
\usepackage{bm}
\usepackage{diagbox}[2011/11/22]

\begin{document}
\title{Guided Training: A Simple Method for Single-channel Speaker Separation}

\author{%
\authorblockN{%
Hao Li, Xueliang Zhang, Guanglai Gao }
\authorblockA{%
{College of Computer Science, Inner Mongolia University, China }
\authorblockA{%
	E-mail:  lihao@mail.imu.edu.cn \  cszxl@imu.edu.cn  \  csggl@imu.edu.cn}}}

\maketitle
\begin{abstract}
Deep learning has shown a great potential for speech separation, especially for speech and non-speech separation. However, it encounters permutation problem for multi-speaker separation where both target and interference are speech. Permutation Invariant training (PIT) was proposed to solve this problem by permuting the order of the multiple speakers. Another way is to use an anchor speech, a short speech of  the target  speaker, to model the speaker identity. In this paper,  we propose a simple strategy to train a long short-term memory (LSTM) model to solve the permutation problem in speaker separation. Specifically, we insert a short speech of target speaker at the beginning of a mixture as guide information. So, the first appearing speaker is defined as the target. Due to the powerful capability on sequence modeling, LSTM can use its memory cells to track and separate target speech from interfering speech. Experimental results show that the proposed training strategy is effective for speaker separation. 
\end{abstract}


\section{Introduction}
\label{sec:intro}
Speech signal for processing and analyzing is usually degraded by interference sources. Separating the target speech from the other interference sources is often referred to as speech separation, which is a challenging but meaningful work.

Over the past decades, several speech separation approaches have been proposed\cite{Xiao2016}\cite{Wang2006}. Speech separation in literature can be broadly decomposed into two categories, multi-channel separation and single-channel separation, depending on the number of the sensors (microphones) applied in the signal recording. Since the multi-channel recording contains more than one sensor, the spatial information for each source can be employed for speech separation, e.g. beamforming technology\cite{Xiao2016}. The effect of multi-channel based methods is directly related to the number of sensors. To achieve a good performance, these methods always need a large number of sensors. However, single channel recording is the most common scenario in the real world. It is also more challenging for target speaker separation. In this paper, we mainly focus on single channel recording.

Recently, deep neural network (DNN) is introduced to solve the signal processing problems and has achieved substantial improvements over the traditional methods\cite{Wang2018}
. In\cite{Delfarah2018}, {Delfarah \textit{et al.}} proposed a RNN-based method for speaker-dependent separation in reverberant environments. More impressively, Deep Clustering (DC)\cite{Hershey2016}, Deep Attractor Networks (DANs)\cite{Luo2018} and PIT\cite{Yu2017} have made speaker-independent separation possible. In particular, DC and DANs combine a neural network and a K-means clustering algorithm cleverly to obtain source separation masks. PIT casts speech separation as a multi-class segregation problem where the supervision is provided as a set instead of an ordered list. DC, DANs and PIT deal with the speech separation as an one-pass problem that separate all sources at once. In addition, they all assume that the correct number of speakers in the test stage is known in advance.


The mixture signal of a target speech with background speech can be separated into individual source signals by the above algorithms. However, these algorithms cannot identify which output signal corresponds to the target speech. This problem is regarded as the permutation problem\cite{Delcroix2018}, and some work has been proposed to solve it by imposing constraints about speaker gender\cite{Wangyannan2017}  or signal intensity \cite{Weng2015}.
In\cite{King2017}, {King \textit{et al.}} proposed EncDec that utilizes an encoder projecting an anchor speech to a fixed-size embedding. The output of the last frame is used as identity of the target speaker, and fed into the decoder to predict the target speech. The encoder and the decoder are jointly trained to improve performance.

Recently, LSTM \cite{Hochreiter1997} shows powerful sequence modeling capabilities on many fields\cite{Graves2013}. In this paper, we make full use of LSTM's capability, and propose a simple training strategy for single-channel target speaker separation. In the method, the first appearing speaker is defined as the target. To ensure that the speech does not appear at the same time, we insert a short duration speech of target speaker at the beginning of a mixture as guide information. Due to LSTM's powerful capability on sequence modeling, the target speaker can be tracked and separated.

The rest of the paper is organized as follows. We will describe our proposed algorithm in Section 2 and Section 3. The experimental setup and evaluation results are presented in Section 4. We conclude this paper in Section 5.
\section{ALGORITHM DESCRIPTION}
\subsection{Problem formulation}
\begin{figure}[htb]
 \centering
\includegraphics[width=0.40\textwidth]{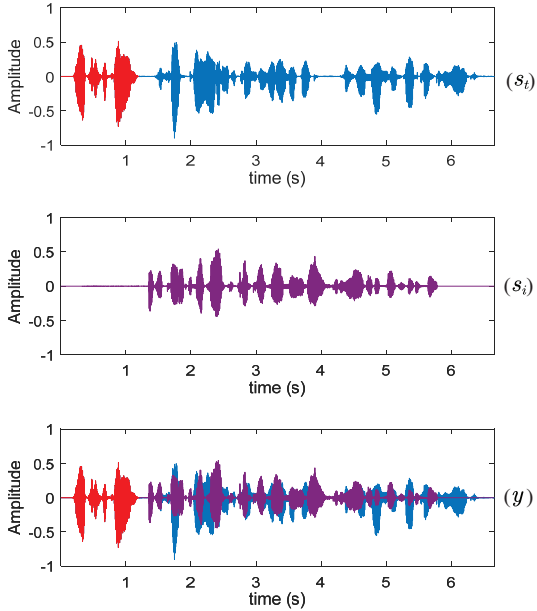}
\caption{Generating mixture speech with interference speaker and target speaker. $s_t$ is the target speaker. The red part in $s_t$ and $y$ is the guided speech. $s_i$ is the interference speaker. $y$ is the mixture.}
\label{fig:mixture_gen}
\end{figure}

The purpose of the method is to extract the first appearing speaker in a linearly mixed signal which can be written as, 
\begin{equation}
y={{s}_{t}}+{{s}_{i}},
\label{eq1}
\end{equation}
 where $y$ is the mixture, $s_t$ and $s_i$ indicate the first speaker and interference speaker signal, respectively. As shown in Fig. \ref{fig:mixture_gen}, the red part in $s_t$ is the guided speech, such as wake-up voice in a smart speaker, which is used to guide the separation model and track the $s_t$ from $y$.
 
The key to speaker tracking is how to use the guide speech segment. RNN seems to be the most suitable neural network model since it makes use of the context information by connecting hidden nodes to the counterparts in the previous step of the sequence. However, gradient vanishing and exploding issues make a vanilla RNN hard to optimize. By introducing a memory cell and employing gate mechanism to control the information flow, LSTM has shown powerful ability to model long range dependencies in the sequential data. Consequently, to capture the long history of the guided speech, LSTM is used to track first speaker directly.
 
\subsection{Features and training target}
Given a time-domain signal with sampling rate being 8000 Hz, The signal is divided into frames by using a 32 ms Hamming window with 16 ms window shift. Fast Fourier transformation (FFT) with 256-point is applied to each frame, which results in 129 frequency bins. In our study, the magnitude spectrum of the mixture is directly used as a feature. To compress the dynamic range of the feature, a cubic root compression is applied. All the features are normalized to zero mean and unit variance by using the statistics of the training data. $\textbf{Y}(m)$ denotes the normalized compressed magnitude feature of mixture at time frame $m$, which is a 129-dimension vector. Then, the proposed system takes the following sequential feature vectors as the input,
\begin{equation}
\textbf{Y}=\left\{ \textbf{Y}\left( 1 \right),\textbf{Y}\left( 2 \right),\ldots ,\textbf{Y}\left( N \right) \right\},
\label{eq2}
\end{equation}
where $N$ is the total number of frames in the utterance. At each time step, the feature of one frame is fed to the system. In other words, no context window is employed.

To get a better reconstruction, the phase sensitive mask (PSM)\cite{Erdogan2015} is used as the training target, which is defined as, 
\begin{equation}
\textbf{M}(n,f) = \frac{{\left| {{\textbf{S}_t}(n,f)} \right|\cos ({\bm{\theta}_y}(n,f) - {\bm{\theta}_{{s_t}}}(n,f))}}{{\left| {\textbf{Y}(n,f)} \right|}}
\label{eq3}
\end{equation}
where $n$ and $f$ are the time frame index and frequency bin index, respectively.  $\bm{\theta}_{y}$ and $\bm{\theta}_{s_t}$ are the phase of mixed speech $\textbf{Y}\left(n,f\right)$ and target speaker $\textbf{S}_{t}\left(n,f\right)$, respectively. The PSM takes phase differences into consideration. The training target can be expressed by the following sequential vectors,
\begin{equation}
\hat{\textbf{M}} = \left\{ {\hat{\textbf{M}}\left( 1 \right),\hat{\textbf{M}}\left( 2 \right), \ldots ,\hat{\textbf{M}}\left( N \right)} \right\}.
\label{eq4}
\end{equation}

Finally, after obtaining the estimation of $\hat{\textbf{M}}$, the time-domain signal is resynthesized by using the phase of mixed speech and ISTFT (inverse STFT), as follows,
\begin{equation}
\begin{split}
\hat{s}_t=\bf{ISTFT}(\hat{M}\circ Y),
\label{eq5}
\end{split}
\end{equation}
the operator $\circ $ denotes the Hadamard product (element-wise product).
\section{NETWORK ARCHITECTURE}
\subsection{LSTM block}
The LSTM block used in this study is defined by the following equations,

\begin{equation}
 \begin{split}
i_t &=sigmoid\left( W_{ii}x_t+W_{hi}h_{t-1}+b_i \right) \\
f_t &=\text{si}gmoid\left( W_{if}x_t+W_{hf}h_{t-1}+b_f \right) \\
g_t &=\tanh \left( W_{ig}x_t+W_{hg}h_{t-1}+b_g \right) \\
o_t &=sigmoid\left( W_{io}x_t+W_{ho}h_{t-1}+b_o \right) \\
c_t &=f_t \circ c_{t-1}+i_i\circ g_t \\
h_t &=o_t \circ  \tanh \left( c_t \right)  ,\\
\label{eq6}
\end{split}
\end{equation}
 \begin{figure*}[ht]
	\centering
	\includegraphics[width=0.50\textwidth]{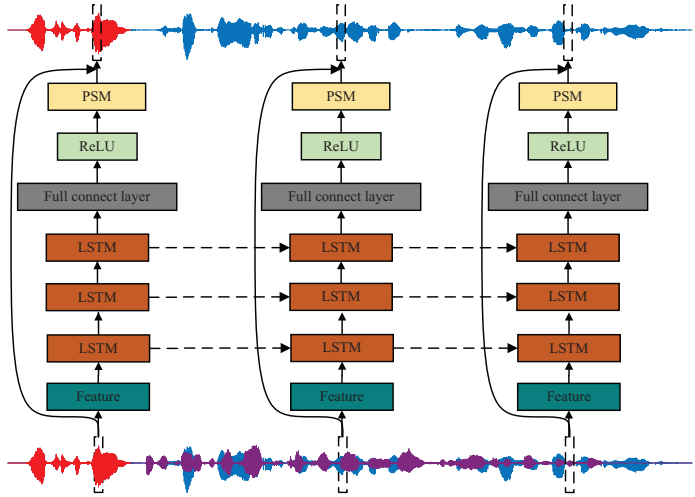}
	\caption{System diagram of the proposed algorithm.}
	\label{fig:propsed_model}
\end{figure*}
where ${{i}_{t}}$, ${{f}_{t}}$, ${{\text{g}}_{t}}$, ${{\text{o}}_{t}}$ are the input, forget, cell and output gates at time step $t$, respectively. ${{h}_{t}}$ is the hidden state. ${{c}_{t}}$ is the memory cell state. ${{x}_{t}}$ is the input of the first layer or the hidden state of the previous layer. $W$ and $b$ denote the weights and biases in the linear transformations, respectively. The subscript $t$ indexes the time step. The initial values are $c_{0}=0$ and $h_{0}=0$.

LSTM introduces the concept of a ``memory cell" with input, output, cell and forget gates, which are also basically recurrent units that have outputs in the range between 0 and 1, and modify the scalars or vectors stored in the cells using the multiplication operation. In the proposed method, the ``memory cell" can store target speaker information for tracing and separating target speaker.
\subsection{System diagram}

The system diagram of the proposed algorithm used in this paper is illustrated in Fig. \ref{fig:propsed_model}, where the feature of the mixture $\textbf{Y}$ is the input, and the PSM is the target. we use three unidirectional LSTM layers followed by one fully-connected layer in the proposed structure. The output layer uses rectified linear units (ReLUs)\cite{Glorot2011} as the activity function to predict the PSM of the first speaker. The number of memory cells in each LSTM is 512. The number of nodes in the fully-connected layer is 1024. The cost function is mean square error (MSE). Weights of the networks are randomly initialized. The ADAM optimizer\cite{Kingma2014} is utilized for back propagation. We also use the dropout\cite{Srivastava2014} in LSTM layers to avoid overfitting. The dropout rate is 0.2.

\section{EXPERIMENTS}
\subsection{Dataset}

The proposed system is evaluated by using the WSJ0-2mix datasets\footnote{Available at: http://www.merl.com/demos/deep-clustering}. In WSJ0-2mix, each sentence contains two speakers. The WSJ0-2mix dataset introduced in\cite{Hershey2016} is derived from the WSJ0 corpus\cite{Garofolo1993}. The 30h training set and the 10h validation set contain two-speaker mixtures generated by randomly selecting from 49 male and 51 female speakers from si\_tr\_s. The Signal-to-Noise Ratios (SNRs) are uniformly chosen between 0 dB and 5 dB. The 5h test set is generated similarly by using utterances from 16 speakers from  si\_et\_05. Which don't appear in the training and validation sets. The test set includes 1603 F\&M sentences, 867 M\&M sentences, and 530 F\&F sentences.

For each mixture, we randomly choose an anchor utterance from the target speaker (different from the utterance in the mixture), and insert the anchor speech to the beginning of the mixture as guide information. The length of the anchor speech is 1 second on average.

\subsection{Metrics and parameters}
The performance is evaluated with two objective metrics: perceptual evaluation of speech quality (PESQ)\cite{Rix2001} and Signal-to-Distortion Ratio (SDR)\cite{Vincent2006}. The PESQ measures the speech quality by computing the disturbance between clean and processed speech. The range of PESQ score is from -0.5 to 4.5. SDR is also a metric widely used to evaluate speech enhancement performance. For both of the PESQ and SDR metrics, the higher number indicates the better performance.
\subsection{Baseline model setting}

We compare the proposed method with EncDec\cite{King2017}. For EncDec, two fully-connected layers with 1024 ReLUs for each one are used in decoder. The encoder uses three unidirectional LSTM layers, the number of memory cells in each LSTM is 512.

\subsection{Evaluation results}

\begin{table}[htb]
\centering
\caption{Average SDR score (dB) on test set.}
\begin{tabular}{|p{1.6cm}|p{0.8cm}|p{0.8cm}|p{0.8cm}|p{1.0cm}|}
\hline
            & \centering F\&M           & \centering F\&F          & \centering M\&M         & \centering Average       \tabularnewline \hline
\centering unprocessed & \centering 2.58           & \centering 2.71          & \centering 2.65         & \centering 2.62          \tabularnewline \hline
\centering EncDec      & \centering 9.10           & \centering 4.47          & \centering 4.55         & \centering 6.97          \tabularnewline \hline
\centering Proposed    & \centering \textbf{9.93} & \centering \textbf{5.84} & \centering \textbf{5.75} & \centering \textbf{8.00} \tabularnewline \hline
\end{tabular}
\label{table:SDR}
\end{table}
\begin{table}[htb]
\centering
\caption{Average PESQ score on test set.}
\begin{tabular}{|p{1.6cm}|p{0.8cm}|p{0.8cm}|p{0.8cm}|p{1.0cm}|}
\hline
            & \centering F\&M           & \centering F\&F          & \centering M\&M         & \centering Average       \tabularnewline \hline
\centering unprocessed & \centering 2.15           & \centering2.13          &\centering 2.25         & \centering 2.17          \tabularnewline \hline
\centering EncDec      & \centering 2.53           & \centering2.22          &\centering \textbf{2.32}         & \centering 2.42          \tabularnewline \hline
\centering Proposed    & \centering\textbf{2.60} & \centering \textbf{2.31} & \centering \textbf{2.32} & \centering \textbf{2.47} \tabularnewline \hline
\end{tabular}
\label{table:PESQ}
\end{table}
\vspace{0em}
 Table \ref{table:SDR} and Table \ref{table:PESQ} show the average scores of SDR and PESQ on the test set, respectively. It can be seen that the proposed method outperforms EncDec. Compared with the mixture, the PESQ score and the SDR of the separated speech improve by 0.30 and 5.38 dB, respectively, indicating the effectiveness of proposed method to perform target speaker separation.

The EncDec extracts the embedding feature using the guided speech in encoder. Then, the fixed embedding serves as an additional input to the decoder. One problem is that, the output of the last frame in encoder does not represent the speaker well \cite{Zmolikova2017}. By taking advantage of LSTM's powerful sequence modeling ability, the proposed method can track the target speaker better, and the guide information stored in memory cells is time-varying according to the mixture. 

It should be mentioned that as the speech of female and male has strong distinguishability\cite{Wangyannan2017}, compared with F\&F (PESQ increased by 0.18, SDR increased by 3.07) and M\&M (PESQ increased by 0.07, SDR increased by 3.10), F\&M (PESQ increased by 0.45, SDR increased by 7.35) has better separation performance.

\begin{table}[htb]
	\centering
	\caption{Results under different time parts.}
	\begin{tabular}{|c|p{0.8cm}|p{0.9cm}|p{0.8cm}|p{0.9cm}|}
		\hline
		\multirow{2}{*}{\centering\diagbox[width=1.9cm, height=0.8cm]{\footnotesize Method}{\footnotesize Criterion}} & \multicolumn{2}{c|}{SDR} & \multicolumn{2}{c|}{PESQ} \\ \cline{2-5} 
		& \centering part I          & \centering part II         & \centering part I          &   \centering part II       \tabularnewline \hline
		unprocessd	& \centering 2.82          & \centering 2.80         & \centering 2.10          &   \centering 2.18       \tabularnewline \hline
		proposed	& \centering 8.17          & \centering 8.19         & \centering 2.43          &   \centering 2.48       \tabularnewline \hline
		$\Delta$	& \centering +5.35          & \centering +5.39         & \centering +0.33          &   \centering +0.30       \tabularnewline \hline
	\end{tabular}
	\label{table:segments}
\end{table}
The performance change during long-term tracking is also explored. Table \ref{table:segments} shows SDR and PESQ results under different time parts, where part I and part II indicate the first and the second half segments of the test speech, respectively. Compared with unprocessed speech, in part I, SDR and PESQ increased by 5.35 dB and 0.33, respectively; in part II, SDR and PESQ increased by 5.39 dB and 0.30, respectively. We can find that the performance on part II is similar to that of the part I, which means that long-term tracking has no effect on performance.
\begin{table}[htb]
	\centering
	\caption{PESQ score for different length of guide speech.}
	\begin{tabular}{|p{1.6cm}|p{0.8cm}|p{0.8cm}|p{0.8cm}|p{1.0cm}|}
		\hline
		& \centering F\&M           & \centering F\&F          & \centering M\&M         & \centering Average       \tabularnewline \hline
		\centering 0.4 s      & \centering 2.53           & \centering 2.26          &\centering 2.28         & \centering 2.41          \tabularnewline \hline
		\centering 0.6 s      & \centering 2.57           & \centering 2.28          &\centering 2.32         & \centering 2.44          \tabularnewline \hline
		\centering 0.8 s      & \centering 2.60 		  & \centering 2.29 		 &\centering 2.33 	      & \centering 2.47  		 \tabularnewline \hline
		\centering 1.0 s      & \centering 2.60           & \centering 2.31          &\centering 2.32         & \centering 2.47          \tabularnewline \hline
		\centering 1.2 s      & \centering 2.61           & \centering 2.30          &\centering 2.32         & \centering 2.47          \tabularnewline \hline
		\centering 1.4 s      & \centering 2.60 		  & \centering 2.32 		 &\centering 2.33 	      & \centering 2.48  		 \tabularnewline \hline
		\centering 1.6 s      & \centering 2.61 		  & \centering 2.32 		 &\centering 2.34 	      & \centering 2.48  		 \tabularnewline \hline
		\centering 1.8 s      & \centering 2.61 		  & \centering 2.31 		 &\centering 2.33 	      & \centering 2.48  		 \tabularnewline \hline
	\end{tabular}
	\label{table:length_pesq}
\end{table}
\begin{table}[htb]
	\centering
	\caption{SDR score (dB) for different length of guide speech.}
	\begin{tabular}{|p{1.6cm}|p{0.8cm}|p{0.8cm}|p{0.8cm}|p{1.0cm}|}
		\hline
		& \centering F\&M           & \centering F\&F          & \centering M\&M         & \centering Average       \tabularnewline \hline
		\centering 0.4 s      & \centering 8.02           & \centering 4.91          &\centering 4.40         & \centering 6.43          \tabularnewline \hline
		\centering 0.6 s      & \centering 9.23           & \centering 5.39          &\centering 5.11         & \centering 7.37          \tabularnewline \hline
		\centering 0.8 s      & \centering 9.82 		  & \centering 5.60 		 &\centering 5.44 	      & \centering 7.81  		 \tabularnewline \hline
		\centering 1.0 s      & \centering 9.93           & \centering 5.84          &\centering 5.75         & \centering 8.00          \tabularnewline \hline
		\centering 1.2 s      & \centering 9.98           & \centering 5.76          &\centering 5.65         & \centering 7.98          \tabularnewline \hline
		\centering 1.4 s      & \centering 9.99 		  & \centering 5.96 		 &\centering 5.78 	      & \centering 8.06  		 \tabularnewline \hline
		\centering 1.6 s      & \centering 10.02 		  & \centering 5.94 		 &\centering 5.79 	      & \centering 8.08  		 \tabularnewline \hline
		\centering 1.8 s      & \centering 10.01 		  & \centering 5.89 		 &\centering 5.78 	      & \centering 8.08  		 \tabularnewline \hline
	\end{tabular}
	\label{table:length_sdr}
\end{table}

We also explore the effect of guide speech length on the results. Table \ref{table:length_pesq} and \ref{table:length_sdr} list the average PESQ and SDR score for different guide speech length on test set. It can be found that the performance becomes better as guide speech duration increases. The reason is that, for a longer guide speech, the model can obtain a better target speaker identify with LSTM memory cells. We also find that the effect will stabilize as the guide speech length reaches approximate 1.4s.

\begin{figure}[htb]
	\centering
	\includegraphics[width=0.27\textwidth]{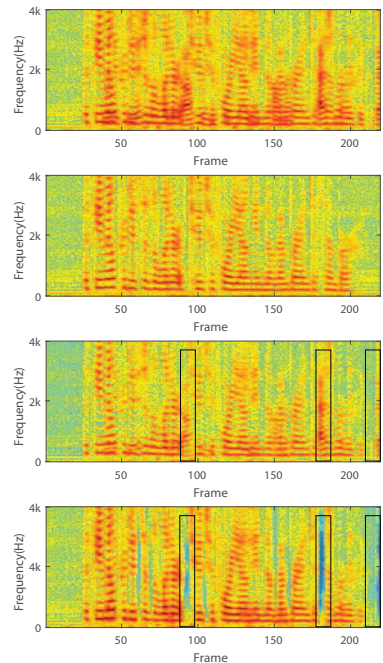}
	\caption{Spectrograms of extracted speech  using different methods. The first graph is the spectrogram of the mixture signal. The second graph is the spectrogram of the target speech. The third graph is the spectrogram of the separated target speech using EncDec method. The fourth graph is the spectrogram of the separated target speech using the proposed method.}
	\label{fig:spec}
\end{figure}

Fig. \ref{fig:spec} illustrates the spectrograms of separated speech using different methods on a test utterance, where both target and interfering speakers are female. Obviously, our method has better performance on target speaker separation. As shown in the black rectangle, the proposed method can suppress the interfering speaker better than EncDec.

\section{CONCLUSION}
In this paper, a simple training strategy for target speaker separation is proposed. By leveraging the capacity of recurrent connections to model the long-term dependencies in speech, the first appearing speaker can be tracked well. According to the experiments results, the proposed method achieves better performance than other baseline methods. In the future, we will explore the robustness problem in the presence of noise and reverberation.
\section{ACKNOWLEDGEMENT}
This research was partly supported by the China National Nature Science Foundation (No. 61876214, No.61773224).
\clearpage
\bibliography{guided_training}
\end{document}